\documentclass[aps,prx,groupedaddress,showpacs,twocolumn]{revtex4-1}
\usepackage{eurosym}
\usepackage[dvipsnames]{xcolor}
\usepackage{hyperref}
\usepackage{graphicx}
\usepackage{amsmath}
\usepackage{amssymb}
\usepackage[export]{adjustbox}

\begin{document}

\title{Space-time quasicrystals in Bose-Einstein condensates}
\author{L. Friedland$^{1}$}
\email{lazar@mail.huji.ac.il}
\author{A. G. Shagalov$^{2}$}
\email{shagalov@imp.uran.ru}
\affiliation{$^{1}$Racah Institute of Physics, The Hebrew University, Jerusalem
91904, Israel.\\
$^{2}$Institute of Metal Physics, Ekaterinburg 620990, Russian
Federation.
}
\begin{abstract}
An autoresonant approach for exciting space-time quasicrystals in
Bose-Einstein condensates is proposed by employing two-component chirped frequency parametric driving or modulation of the interaction strength within Gross-Pitaevskii equation. A weakly nonlinear theory of the process is
developed using Whitham's
averaged variational principle yielding reduction to a two-degrees-of-freedom dynamical system in action-angle variables. Additionally, the theory also delineates permissible driving parameters and establishes thresholds on the driving amplitudes required for autoresonant excitation.
\end{abstract}
\maketitle
\section{Introduction}
Quasicrystals in materials are structures ordered in space, but not exactly
periodic \cite{Shechtman,Levine,Senechal}. By analogy, time or space-time quasicrystals are ordered
systems in space-time, but not exactly periodic in space and/or time. By now,
these systems have been studied in periodically driven magnon condensates \cite{Autti,Kreil}
and ultracold atoms \cite{Giergiel,Cosme}, where temporal symmetry was destroyed due to
subharmonic response. In this study, we explore a different path to
space-time quasicrystals. It is well known that a number of so-called \textit{integrable} nonlinear partial differential equations (PDE's) have \textit{multiphase} solutions \cite{Scott} of the form $f(\theta
_{1},\theta _{2},...,\theta _{N};\lambda _{1},\lambda _{2},...,\lambda _{N})$, where $N$ phase variables $\theta _{i}=k_{i}x-\omega _{i}t$ have wave numbers 
$k_{i}$ which are multiples of some $k_{0}$ (the solution is spatially
periodic), $\lambda _{i}$ are constants, while frequencies $\omega
_{i}$ are functions of $k_{i}$ and $\lambda _{i}$. By choosing some set of $\lambda _{i}$, one can make some or all of these frequencies incommensurate,
forming an ideal space-time quasicrystalline structure, which is periodic in
space and aperiodic in time, still having a complex long range time ordering.
Examples of such PDEs are the Korteweg-de-Vries (KdV), sine-Gordon (SG), and nonlinear Schrodinger (NLS) equations, which find multiple
application in physics \cite{Scott}. However, why these ideal space-time quasicrystals are
not yet realized in experiments? The answer lies in complexity, as their
analysis typically requires advanced mathematical methods, such as the
inverse scattering transform (IST)  \cite{Novikov}, while experimental realization depends on
forming complicated space-time dependent initial conditions, which is an
unrealistic task.

In this work, we suggest a different approach to realizing space-time quasicrystals based on autoresonance  \cite{LazarWiki}. Autoresonance is an important nonlinear phenomenon, where a system phase-locks to a 
\textit{chirped} frequency driving perturbation and remains phase-locked
continuously for an extended period of time, despite variations of the
driving frequency. As the driving frequency varies in time, so does the
frequency of the excited solution, leading to formation of a stable highly
nonlinear state. We will focus on autoresonant formation and
control of \textit{two-phase} excitations of the Gross-Pitaevskii (GP) equation
describing Bose-Einstein condensates (BECs). The excitation proceeds from \textit{trivial} initial conditions and uses a combination of two independent \textit{small} amplitude wave-like driving perturbations. In contrast to existing applications using
large amplitude pulsed drives or optical crystalline structures for
containing BEC's, we can remove the driving perturbation after some time,
remaining with a free, slightly perturbed, but stable nonlinear two-phase GP
solution. These excited quasicrystalline structures are controlled by two independently
chirped driving frequencies, thus exploring a continuous range of
parameters $\lambda _{i}$, i.e., a continuous set of space-time quasicrystals. The autoresonance approach has been used previously in excitation
of multi-phase waves in different applications with the theory based on the
IST method \cite{Lazar2003,Lazar2005,Lazar2009}. Here we will apply a \textit{simpler analysis} of the
process of capture into a double autoresonance in the system using two
chirped frequency drives, similar to recent studies on the formation of two-phase
waves in plasmas \cite{Munirov1,Munirov2}.

Our presentation will be as follows. In the next section, we illustrate the
formation of space-time quasicrystals in BECs through numerical simulations.
Section III presents the quasi-linear theory of formation of GP
quasicrystals using Whitham's averaged variational approach \cite{Whitham}. Section IV
addresses the problem of the allowed parameter space for autoresonant
excitations and with the associated threshold phenomenon on
the driving amplitudes. Finally, Sec. V presents our conclusions.

\section{Quasicrystals in a BEC via simulations}
The basic model for studying nonlinear dynamics of BECs is GP equation \cite{Dalfovo},written in dimensionless form 
\begin{equation}
i\varphi _{t}+\varphi _{xx}-\mathbf{U}(x,t)\varphi +g(x,t)|\varphi
|^{2}\varphi =0.  \label{GP}
\end{equation}%
Here, time is measured in units of the inverse transverse trapping frequency $%
\omega _{\bot }^{-1}$, while space and density are measure in units of $l_{\bot }=[\hbar
/(2m\omega _{\bot })]^{1/2}$ and $m\omega _{\bot }/2\pi \hbar |a_{0}|$,
respectively, where $m$ represents the atomic mass. In Eq. (1), $g=2a(x,t)/|a_{0}|$
is the normalized, space-time modulated interaction strength, where $a$ represents
the $s$-wave scattering length of interacting particles in the BEC. For
condensates with repulsive interactions of particles $a<0$ and for attractive interactions $a>0$. $\mathbf{U}(x,t)$ denotes the longitudinal
potential. We will assume that our system is perturbed by a combination of
independent, small amplitude waves 
\begin{equation}
f=\varepsilon _{1}\cos [k_{1}x-\psi _{1}(t)]-\varepsilon _{2}\cos
[k_{2}x-\psi _{2}(t)],  \label{perturb}
\end{equation}%
where $\psi _{i}(t)=\int \omega _{di}(t)dt$ and $\omega _{di}(t)=\omega _{0i}-\alpha _{i}t$ are slowly chirped driving
frequencies.  We consider two
driving options. The first is a parametric-type driving $\mathbf{U}(x,t)=f$, $|f|\ll1$, while
the interaction strength is not perturbed, i.e., $g=2\sigma $ and $\sigma
=\pm 1$. The second driving scenario is a modulation of the interaction strength by external magnetic field,for example,
near Feshbach resonance \cite{Yamazaki}. In this case, we assume $\mathbf{U}=0$,
\begin{equation}
g(x,t)=2\sigma (1+f),  \label{interaction}
\end{equation} and again $|f|\ll1$.
For both driving options we can rewrite Eq. (\ref{GP}) as a weakly perturbed
NLS equation%
\begin{equation}
i\varphi _{t}+\varphi _{xx}+2\sigma |\varphi |^{2}\varphi =F\varphi ,
\label{NLS}
\end{equation}%
where $F\ $is either $-f$ or $-2\sigma |\varphi |^{2}f$. In computer
simulations below, we will use the parametric driving, assume periodic boundary conditions $\varphi (x,t)=\varphi (x+l,t)$, thus $k_{1,2}$ are multiples of $k_{0}=2\pi /l$. Nevertheless,
in Secs. III and IV, we will discuss both driving options. It
should be mentioned that periodic boundary conditions are usually
assumed in numerical simulations of an infinite domain for a spatially
periodic driving as well as in 1D modeling along the torus-like BEC
(ring-trap geometry) \cite{Schloss,Zhu}. A special case when the driving \eqref{interaction} is a standing wave, i.e. $\varepsilon_{1}=\varepsilon_{2}$, $\psi_{1}=\psi_{2}$, and $k_1=-k_2$ was studied recently \cite{solitons}, so the present investigation is a generalization to two \textit{independent} driving components.

In the periodic case, the unperturbed ground state of a BEC is a spatially
homogeneous solution of Eq. 
 \eqref{NLS}
\begin{equation}
\varphi (x,t)=U_{0}e^{2i\sigma U_{0}^{2}t}  \label{start}
\end{equation}%
with constant amplitude $|\varphi |=U_{0}$. The frequency of a perturbation of the
homogeneous state is \cite{Leggett}
\begin{equation}
\omega _{0}=k\sqrt{k^{2}-4\sigma U_{0}^{2}}.  \label{fres}
\end{equation}%
\begin{figure}[tp]
\includegraphics[width=0.51\textwidth, height=0.28\textheight, left]{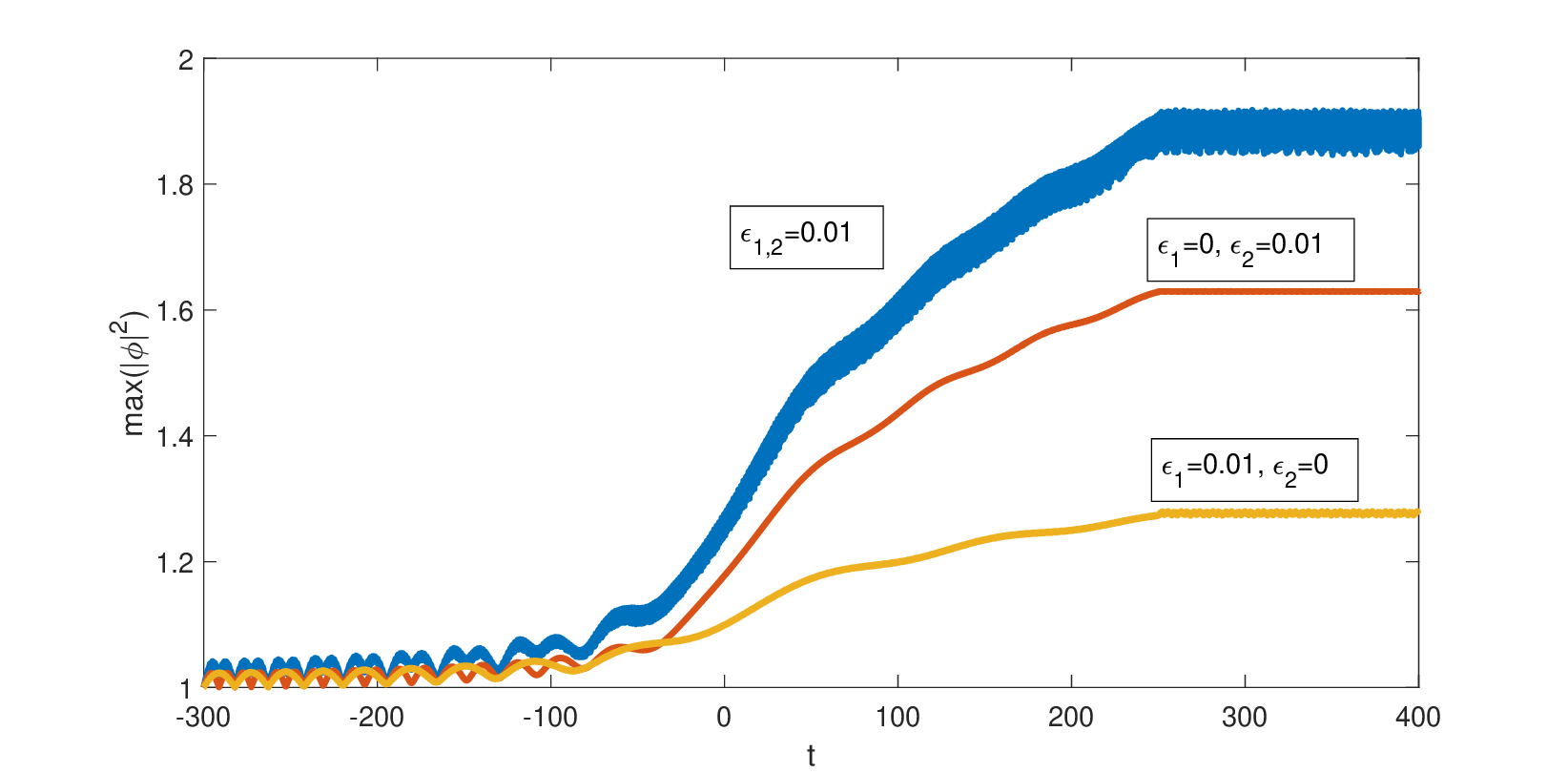}
\caption{Maximum of $|\varphi|^2$ over x versus time. The upper line corresponds to nonzero driving amplitudes $\epsilon _{1,2}$. The middle and the lower lines describe the cases when either $\epsilon _{1}=0$ or $\epsilon _{2}=0$, respectively. In all cases the drives are switched off at $t=250$}
\label{fig1}
\end{figure}
Condensates with repulsive interaction of particles, when $\sigma =-1$, are
stable. In this case, frequency (\ref{fres}) is known as the \textit{Bogolubov}
frequency. Dark solitons are typical structures in these condensates. In
the opposite case ($\sigma =1$), bright solitons exists. In this case $\omega
_{0}$ can be imaginary, leading to modulational instability. This
instability is well-known in plasma physics and nonlinear optics \cite%
{Zakharov,Agrawal}. If a condensate has a length $l$, then the wave number of
the main mode is $k_{0}=2\pi /l$ and the stability condition restricts
the density of the condensate to $U_{0}^{2}<\pi ^{2}/l^{2}$. If the condensate
has a cigar-like shape with the transverse dimension $l_{\bot }$, then, in
physical variables, the stability condition can be written as a
restriction on the number of particles, $n<(l_{\bot }/l)(l_{\bot }/|a_{0}|).$

\begin{figure}[tp]
\includegraphics[width=0.51\textwidth, height=0.31\textheight, left]{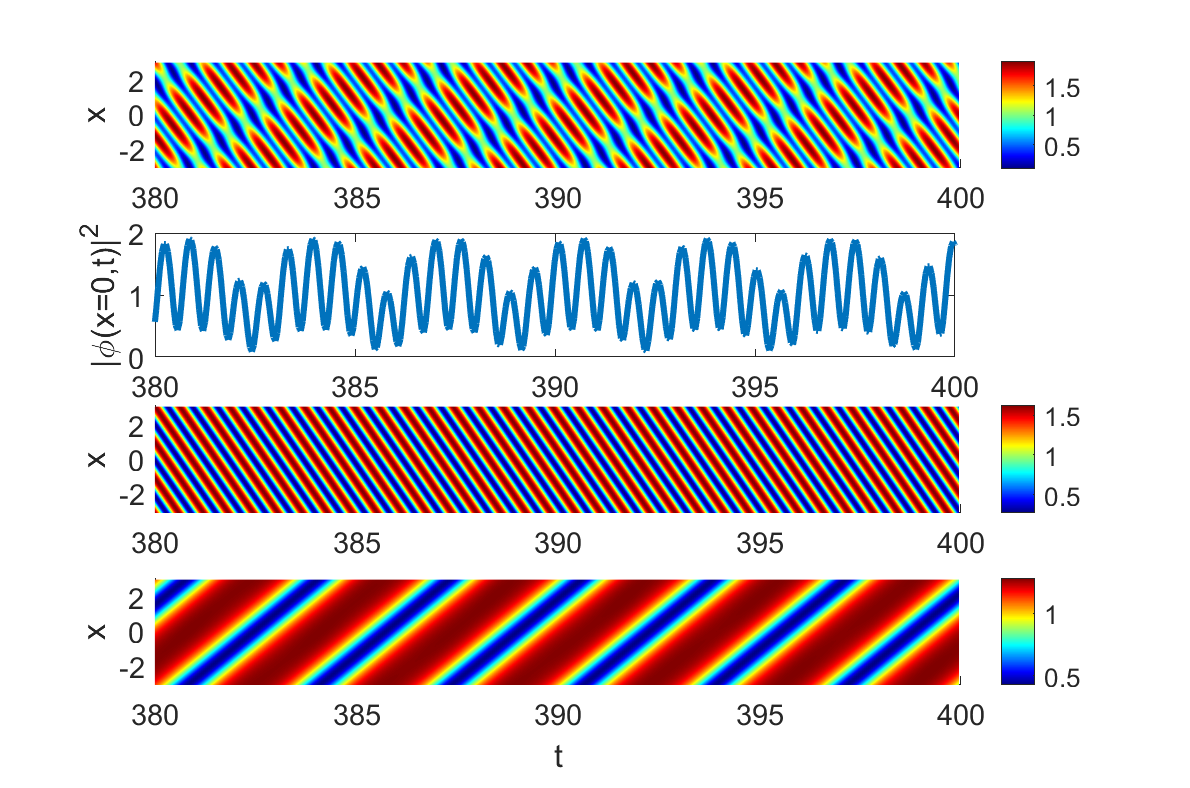}
\caption{Two- and one-phase autoresonant excitations in simulations. The upper panel shows the color map of a two-phase solution $|\varphi|^2$ in space-time for driving parameters $\varepsilon _{1,2}=0.01$, $k_{1,2}=1,-3$, chirp rates $\alpha _{1,2}=0.0012,0.0024$. The second panel from the top shows $|\varphi|^2$ versus time at $x=0$, illustrating 1:5 quasi-periodicity of the solution. The lowest two panels show color maps of autoresonant single-phase excitations $|\varphi|^2$ for the same parameters and initial conditions as in the upper panel, but when only one of the diving components is applied.}
\label{fig2}
\end{figure}
We proceed to numerical simulations of autoresonant formation of a 
space-time GP quasicrystal by focusing on the case of $\sigma =-1$ and starting
in the ground state (\ref{start}) with $U_{0}=1$. The driving parameters are $\varepsilon _{1,2}=0.01$, $k_{1,2}=1,-3$, chirp rates $\alpha _{1,2}=0.0012,0.0024$, and $\omega _{0i}$
are given by Eq. \eqref{fres} for each $k_{i}$.
The simulation begins at $t_0=-300$ and both components of the drive pass the corresponding Bogolubov resonances at $t=0.$ Furthermore, we
switch off the drives at $t=t_{s}=250$. The
corresponding time dependence of the maximum (over $x$) $|\varphi
|^{2}$ is shown in Fig. 1 by the upper blue line. One can observe that the
excitation amplitude increases continuously until the drive is turned off, and the maximal amplitude
remains nearly stationary afterwards. We also show the excited quasicrystalline
structure in the time interval $380<t<400$ (after the driving is switched off) using a
colormap in the upper panel of Fig. 2. Furthermore, the ratio of the two driving frequencies at 
$t=t_{s}$ in this example is approximately $1:5$. The second panel from the top in Fig. 2 shows $|\varphi|^2$ versus time in the same example at $x=0$. One can see the short and long driving periods in the panel illustrating $1:5$ quasi-periodicity and the two-phase locking with the drives.

To further
illustrate the characteristics of autoresonant excitation in the system, we
show, by the lower two lines in Fig. 1, the cases when only one of the two drives is present in the same example and illustrate the colormaps of the associated excitations in the lowest two panels in Fig. 2. A single-phase parametric autoresonant excitation in
this system was analyzed in Ref. \cite{Lazar2022}. In this case, one forms a growing
amplitude nonlinear wave traveling with the phase velocity $\omega
_{di}/k_{i}$ of the corresponding drive. The directions of these propagation
velocities are clearly seen in Fig. 2. We observe the
same two characteristic directions in the upper panel in Fig. 2 corresponding to the two-phase autoresonant excitation, illustrating again the continuing phase locking
with the two driving components. 

One of the most important issues associated with the autoresonance is the threshold on the driving amplitudes for the continuing
phase locking in the system.
Figures 1 and 2 show that the two-phase
autoresonant excitation (where both drives are present) is very different from
that with a single drive. This means that the driving amplitudes must be
sufficiently large to obtain a two-phase quasicrystalline structure, leading to the problem of thresholds for the transition to autoresonance. We illustrate this sudden transition in Fig. 3, showing the maximal $|\varphi|^2$ versus time in three cases with the  same parameters as in Fig. 1, but $\varepsilon _{2}=0.006,0.005$, and $0.004$, while keeping $\epsilon_1=0.01$. One observes a sharp transition when passing from $\epsilon_2=0.005$ to $0.004$. The color maps of $|\varphi|^2$ in space-time in these three cases are shown in  Fig. 4. One can see that the quasicrystalilne structure in the upper two panels in the Figure is similar to that shown in the upper panel in Fig. 2, but the structure changes significantly as one passes to $\epsilon_2=0.004$. The lowest panel in Fig. 4 has smaller amplitude and is closer to the lowest panel in Fig. 2, corresponding to the single phase excitation. We interpret this transition as the loss of the phase locking when $\epsilon_2$ is below some threshold value between $\epsilon_2=0.004$ and $0.005$.
\begin{figure}[tp]
\includegraphics[width=0.51\textwidth, height=0.28\textheight, left]{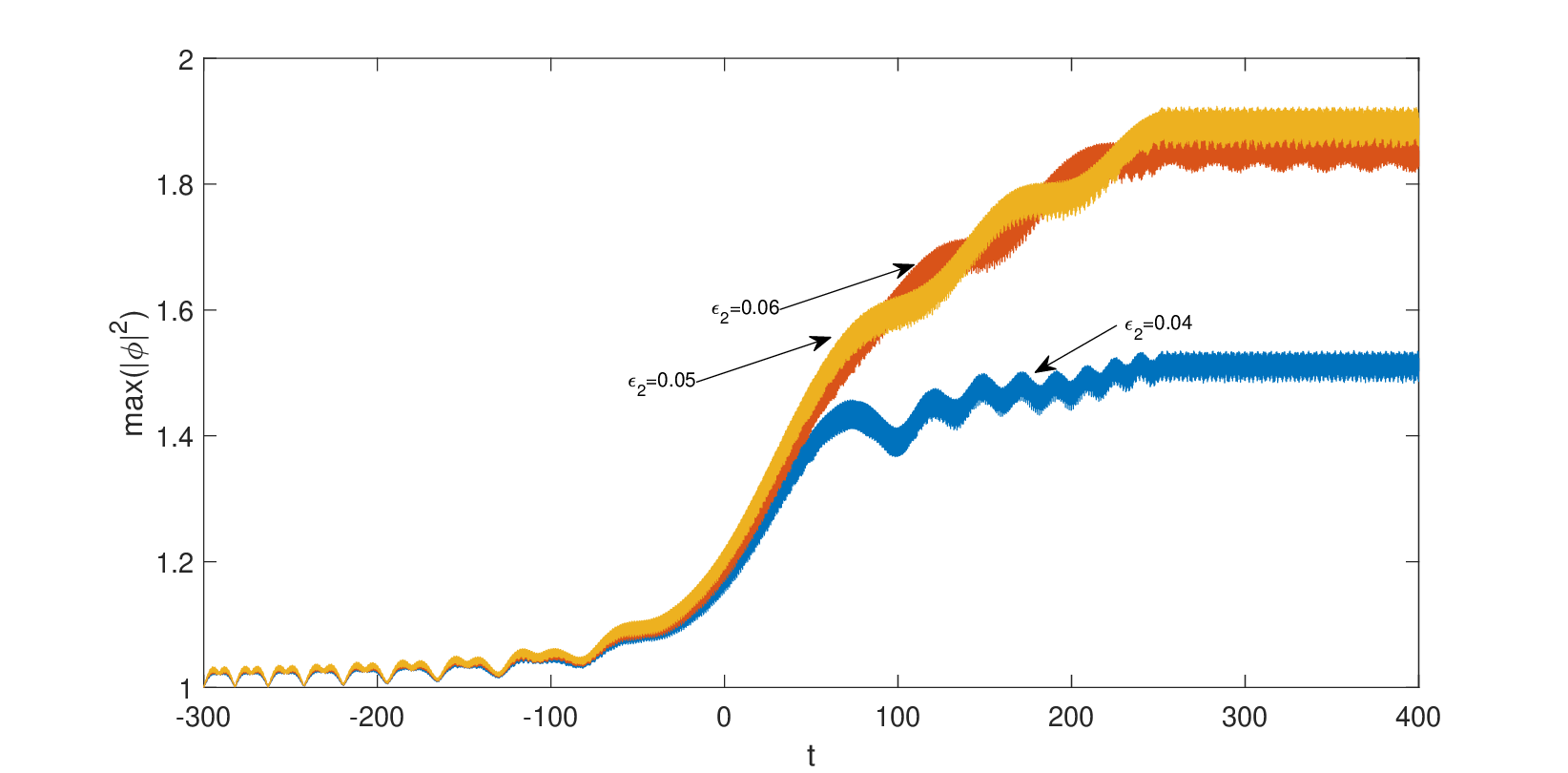}
\caption{The passage through threshold. The figure shows the color maps of $max(|\varphi|^2)$ over $x$ versus time for three values of $\epsilon_2=0.006,0.005,0.004$. In all cases $\epsilon_1=0.01$, $\alpha_1=0.0012$ and $\alpha_2=0.0024$.}
\label{vel}
\end{figure}
\begin{figure}[bp]
\includegraphics[width=0.51\textwidth, height=0.28\textheight, left]{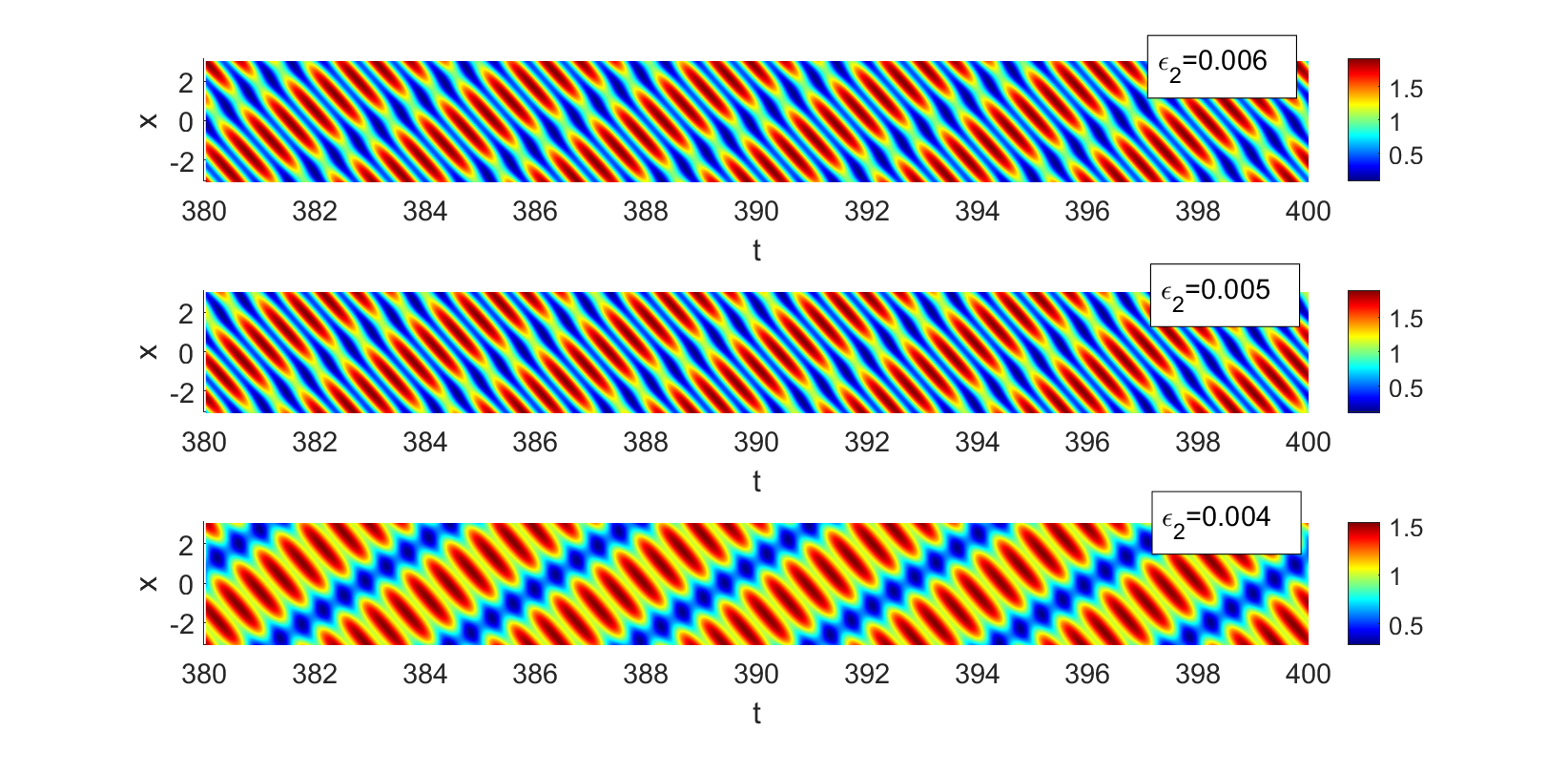}
\caption{The passage through threshold. The colormaps of $|\varphi|^2$ over $x$ in space-time in three examples shown in Fig. 3.}
\label{vel}
\end{figure}

As in all single
phase autoresonant interactions, the double autoresonant phase locking in the driven GP system
starts in the initial excitation stage, as the two drives simultaneously pass
through the linear (Bogolubov) frequencies in the problem. We also find that the autoresonant phase locking is a \textit{weakly nonlinear} phenomenon. Therefore, the
next section is devoted to the quasi-linear theory of two-phase GP
autoresonance.\section{Weakly nonlinear theory}

In developing the weakly nonlinear theory of autoresonant two-phase solutions of the GP equation, we will focus primarily on the parametric-type driving scenario. The case of
driving by modulation of the interaction strength is obtained similarly and
will be briefly discussed at the end of this Section. Therefore, we proceed from Eq.
(4) for the ponderomotive case 
\begin{equation}
i\varphi _{t}+\varphi _{xx}+2\sigma \left\vert \varphi \right\vert
^{2}\varphi =-\varphi (\varepsilon _{1}\cos \theta _{d1}-\varepsilon
_{2}\cos \theta _{d2}),  \label{101}
\end{equation}%
where $\theta _{di}=k_{i}x-\psi _{di}(t)$ and seek solution of form $\varphi
=U\exp (iV)$ governed by the following set of real equations 
\begin{equation}
U_{t}+V_{xx}U+2V_{x}U_{x}=0,  \label{102}
\end{equation}
\begin{equation}
V_{t}U-U_{xx}+V_{x}^{2}U-2\sigma U^{3}=U(\varepsilon _{1}\cos \theta
_{d1}-\varepsilon _{2}\cos \theta _{d2}).  \label{103}
\end{equation}%
The Lagrangian density for this problem is 
\begin{eqnarray} 
L=&&\frac{1}{2}\left[ U_{x}^{2}+U^{2}(V_{x}^{2}+V_{t})\right]-\frac{\sigma }{2%
}U^{4} \nonumber\\
&&-\frac{U^{2}}{2}(\varepsilon _{1}\cos \theta _{d1}-\varepsilon
_{2}\cos \theta _{d2}).  \label{104}
\end{eqnarray}%
This Lagrangian representation suggests using Whitham's averaged variational
approach \cite{Whitham} to analyze our problem. The first step in this
direction is to assume constant frequency drives, $\psi _{di}=\omega _{di}t$%
, and seek \textit{phase-locked} solutions of the linearized problem of form 
\begin{eqnarray}
U &=&U_{0}+U_{1}\cos \theta _{d1}-U_{2}\cos \theta _{d2},  \label{105} \\
V &=&2\sigma U_{0}^{2}t+V_{1}\sin \theta _{d1}+V_{2}\sin \theta _{d2},
\label{106}
\end{eqnarray}%
[Note that our unperturbed solution is $\varphi _{0}=U_{0}\exp (2i\sigma
U_{0}^{2}t)$]. Then, by linearization, and neglecting products of linear
amplitudes $U_{i}$ and $\varepsilon _{i}$, Eqs (\ref{102}) and (\ref{103})
become
\begin{equation}
\omega _{di}U_{i}-k_{i}^{2}U_{0}V_{i}=0,  \label{107}
\end{equation}%
\begin{equation}
-\omega _{di}V_{i}U_{0}+(k_{i}^{2}-4\sigma U_{0}^{2})U_{i}=\varepsilon
_{i}U_{0},  \label{108}
\end{equation}%
yielding solutions%
\begin{eqnarray}
V_{i} &=&\frac{\varepsilon \omega _{di}}{(\omega _{0i}^{2}-\omega _{di}^{2})}%
,  \label{109} \\
U_{i} &=&\frac{k_{i}^{2}U_{0}}{\omega _{di}}V_{i},  \label{110}
\end{eqnarray}%
where the linear resonance frequencies 
\begin{equation}
\omega _{0i}=\left\vert k_{i}\right\vert \sqrt{k_{i}^{2}-4\sigma U_{0}^{2}}.
\label{110a}
\end{equation}

Now, we proceed to chirped-driven problem, where $\psi _{i}=\int \omega
_{di}(t)dt$ and extend Eqs. (\ref{105}) and (\ref{106}) to next nonlinear
order
\begin{eqnarray}
&U&=U_{0}+U_{1}\cos \theta _{1}-U_{2}\cos \theta _{2}+u_{0}+u_{11}\cos
(2\theta _{1}) \label{111}\\
&&+u_{22}\cos (2\theta _{2})+u_{12p}\cos (\theta _{1}+\theta
_{2})+u_{12m}\cos (\theta _{1}-\theta _{2}),  \nonumber
\end{eqnarray}
\begin{eqnarray}   
&V&=2\sigma U_{0}^{2}t+V_{1}\sin \theta _{1}+V_{2}\sin \theta _{2}+\xi+v_{11}\sin (2\theta _{1}) \label{112} \\
&&+v_{22}\sin (2\theta _{2})+v_{12p}\sin (\theta
_{1}+\theta _{2})+v_{12m}\sin (\theta _{1}-\theta _{2}). 
\nonumber
\end{eqnarray}%
Here $U_{i}$ and $V_{i}$ are small (viewed as first-order perturbations),
while all other amplitudes are assumed to be of second-order
in $U_{i}$ and $V_{i}$. In these solutions $\theta _{i}=k_{i}x-\psi _{i}$
and $\psi _{i}=\int \omega _{i}(t)dt$ is a new fast \textit{independent}
variable. At this stage, we do not assume phase-locking in the
system, but view the difference $\Phi _{i}(t)=\psi _{i}-\psi _{di}$ as
slow function of time. Similarly, all the amplitudes are also assumed to be slowly
varying functions of time. The reason for choosing the second- order ansatz
of this form is consistent with the form of the Lagrangian density
containing either different powers of $U$ or products of derivatives of $V$
and powers of $U$. The auxiliary phase $\xi=\int \gamma(t)dt$ in
Eq. (\ref{112}) is necessary because $V$ is the potential (it enters the
Lagrangian density via derivatives only \cite{Whitham}).

The next step is to replace $\theta _{di}=\theta _{i}+\Phi _{i}(t)$ in the
driving part of the Lagrangian density (\ref{104}), substitute the above ansatz into the
Lagrangian density, and average it over $\theta _{i}\in \lbrack
0,2\pi ].$ This averaging is done via the Mathematica package in the Appendix. The resulting averaged Lagrangian
density $\Lambda=\Lambda
(U_{1,2},V_{1,2},u_{0},u_{11},u_{22},u_{12p},u_{12m},v_{11},v_{22},v_{12p},v_{12m},$ $\Phi _{1,2},\xi)$
 is a function of all $13$ slow first and second-order amplitudes and $\Phi
_{1,2}$ and $\xi$. The Lagrange's equations for
all these $16$ variables, form a system describing slow
autoresonant evolution in the problem. Reducing this problem to a smaller
set of evolution equations involves tedious algebra, which, nevertheless,
can be performed using the Mathematica package. The details of this
reduction are given in the Appendix, and here we present the final closed
system of $4$ equations for $U_{1},U_{2},\Phi _{1},$ and $\Phi _{2}$ (see
Eqs.(\eqref{A23},\eqref{A24},\eqref{A26}) in the Appendix):

\begin{equation}
\frac{dU_{i}}{dt}=-\frac{\varepsilon _{i}U_{0}k_{i}^{2}}{2\omega _{0i}}\sin
\Phi _{i}.  \label{113a}
\end{equation}

\begin{eqnarray}
(\omega _{01}^{2}-\omega
_{1}^{2})U_{1}&-&24U_{0}^{2}U_{1}^{3}+4k_{1}^{2}\sigma
U_{1}(2U_{1}^{2}+U_{2}^{2})\nonumber\\
&&-\varepsilon _{1}U_{0}k_{1}^{2}\cos \Phi _{1}=0.  \label{113} 
\end{eqnarray}
\begin{eqnarray}
(\omega _{02}^{2}-\omega
_{2}^{2})U_{2}&-&24U_{0}^{2}U_{2}^{3}+4k_{2}^{2}\sigma
U_{2}(2U_{2}^{2}+U_{1}^{2})\nonumber\\
&&-\varepsilon _{2}U_{0}k_{2}^{2}\cos \Phi _{2}=0.  \label{114}
\end{eqnarray}
\begin{figure}[bp]
\includegraphics[width=0.49\textwidth, height=0.28\textheight, left]{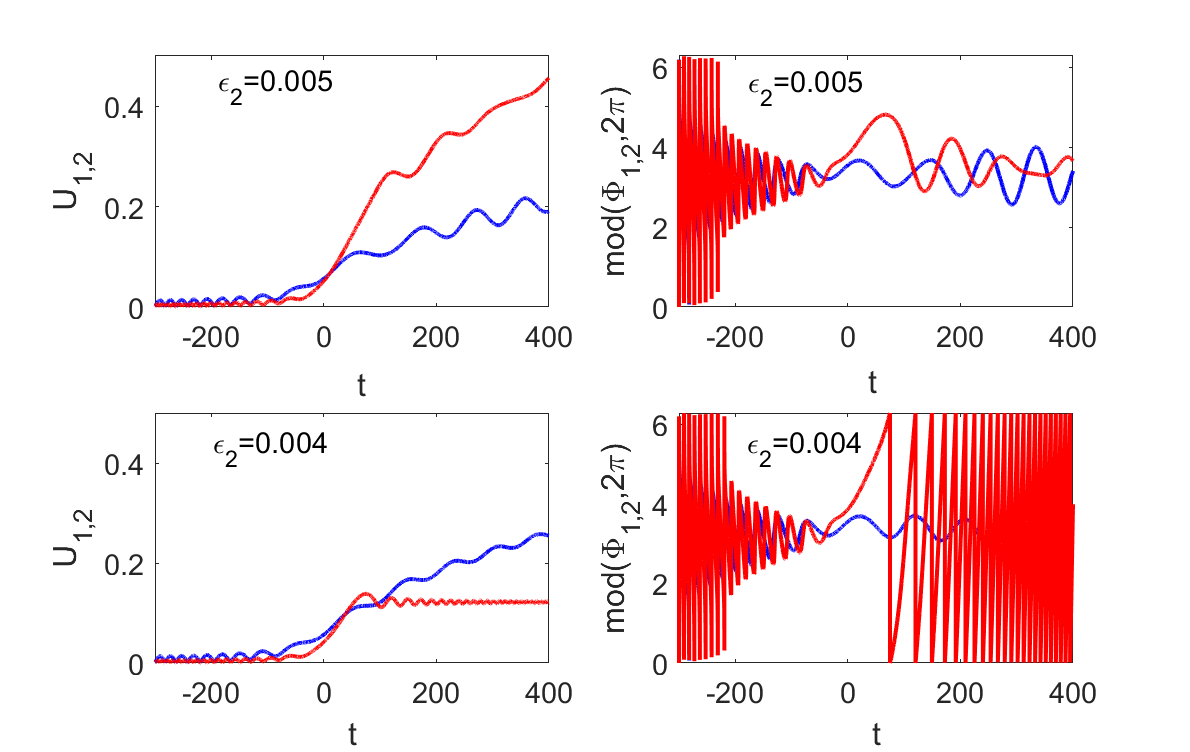}
\caption{Threshold phenomenon. The upper two panels illustrate double phase-locking in the system for $\epsilon_2=0.005$, while in the lower two panels, where $\epsilon_2=0.004$, the phase locking with one of the driving components is lost. All other driving parameters in the two cases are the same as in Fig. 4, i.e., $\epsilon_{1}=0.01,\alpha_{1,2}=0.0012,0.0024$.}
\label{fig5}
\end{figure}
All remaining dependent variables in the problem, i.e., $
V_{1,2}$, $u_{0},u_{11},u_{22},u_{12p},u_{12m},v_{11},v_{22},v_{12p},v_{12m}$, $\Phi _{1,2}$, $\xi$ are related to $U_{1},U_{2}$ (see Eqs. (\ref{A11}-\ref{A20}) in
the Appendix). Before proceeding to the analysis of this system, we rewrite
Eqs. (\eqref{113},\eqref{114}) explicitly as differential equations for $\Phi
_{1,2}$. We approximate $\omega _{0i}^{2}-\omega _{i}^{2}\approx
2\omega _{0i}(\omega _{0i}-\omega _{i})=2\omega _{0i}(\omega _{di}-\omega
_{i}+\alpha _{i}t)=-2\omega _{0i}(\frac{d\Phi _{i}}{dt}-\alpha _{i}t)$,
which allows to write these equations as%
\begin{eqnarray}
\frac{d\Phi _{1}}{dt} &=&\alpha _{1}t+\frac{4\sigma (-3\sigma
U_{0}^{2}+k_{1}^{2})}{\omega _{01}}U_{1}^{2}+\frac{2k_{1}^{2}\sigma }{%
\omega _{01}}U_{2}^{2} \nonumber\\&-&\frac{\varepsilon _{1}U_{0}k_{1}^{2}}{2\omega
_{01}U_{1}}\cos \Phi _{1},  \label{115a} 
\end{eqnarray}
\begin{eqnarray}
\frac{d\Phi _{2}}{dt} &=&\alpha _{2}t+\frac{4\sigma (-3\sigma
U_{0}^{2}+k_{2}^{2})}{\omega _{02}}U_{2}^{2}+\frac{2k_{2}^{2}\sigma }{\omega
_{02}}U_{1}^{2} \nonumber\\
&-&\frac{\varepsilon _{2}U_{0}k_{2}^{2}}{2\omega _{02}U_{2}}%
\cos \Phi _{2}.  \label{116}
\end{eqnarray}%

Equations \eqref{113a},\eqref{115a}, and \eqref{116} comprise a complete set of differential\
equations for studying the dynamics in our problem. By solving this system,
and calculating all second-order objects as described above, we obtain a full
quasilinear two-phase solution (\ref{111},\ref{112}) of the chirped-driven GP equation. As
an illustration, Fig. 5 shows the dynamics of $U_{1,2}$ and $\Phi _{1,2}$ for two
cases with the same parameters as in the two lower panels in Fig. 4. The figure illustrates the loss of phase-locking with one of the components of the drive as one passes from $\epsilon_2=0.005$ to $0.004$. The upper two panels in Fig. 5 show double phase locking of $\Phi_{1,2}$ at $\pi$ [$mod(2\pi)$] and a continuous autoresonant growth of $U_{1,2}$, while in the lower two panels only $U_1$ continues to grow, while $U_2$ saturates as $\Phi_2$ escapes. Thus, one requires both driving amplitudes to be above some minimal values for a persisting double autoresonance in the system. We discuss this threshold effect in the next section.
\begin{figure}[tp]
\includegraphics[width=0.51\textwidth, height=0.28\textheight, left]{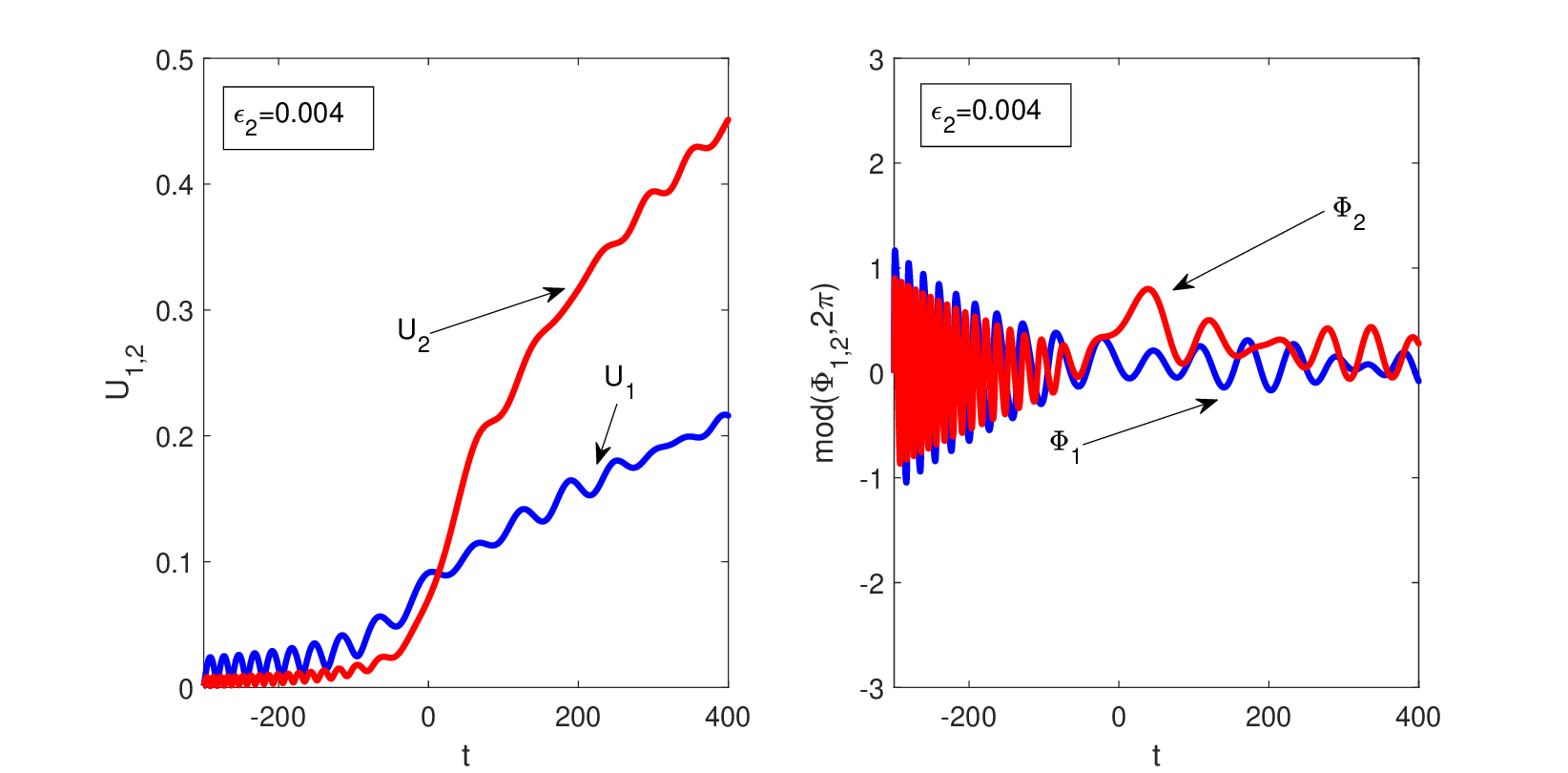}
\caption{Double phase-locking in the system driven by modulation of the interaction strength. The parameters are the same as in the lower two panels in Fig. 5, but $U_0$ in the driving term is replaced by $2\sigma U_0^3$.}
\label{fig6}
\end{figure}

We conclude this section by discussing the case of driving via the modulation of the interaction strength (see the comments at the beginning of Sec. II). It is shown at the end of the Appendix that the reduced system of equations describing two-phase autoresonant formation of space time GP quasicrystals in this case is the same as for the ponderomotive drive (see Eqs. \eqref{113a},\eqref{115a},\eqref{116}), but $U_0$ in the driving terms must be replaced by $2\sigma U_0^3$. We present an example of such a case in Fig. 6, where the parameters are the same as in the lower two panels in Fig. 5, but the driving is via the interaction strength in the GP equation, as discussed above. One can see two differences between between Fig. 5 and 6. One is that in the case of the interaction strength drive the double phase-locking is restored. This is because effectively the driving amplitude increased by a factor of two, and passed the threshold. The second difference is that the phase mismatches $\Phi_{1,2}$ in Fig. 6 are locked near $0$. The reason is that in this case, there is a new factor of $\sigma$ in the driving perturbation, which for $\sigma=-1$ in this example changes the phase-locking location, as will be discussed in the next Section.
\section{Autoresonance conditions and threshold phenomenon}
In this section, we discuss conditions for the autoresonant evolution of 
space-time quasicrystals in BECs. We proceed by defining new (action)
variables%
\begin{equation}
I_{1,2}=\frac{\omega _{r1,2}U_{1,2}^{2}}{k_{1,2}^{2}}.  \label{201}
\end{equation}%
Then the system of Eqs. (\ref{113a},\ref{115a},\ref{116}) describing two-phase BECs can
be rewritten as
\begin{eqnarray}
\frac{dI_{1}}{dt} &=&-\eta _{1}\sqrt{I_{1}}\sin \Phi _{1},  \label{202} \\
\frac{dI_{2}}{dt} &=&-\eta _{2}\sqrt{I_{2}}\sin \Phi _{2},  \label{203}
\end{eqnarray}%
\begin{eqnarray}
\frac{d\Phi _{1}}{dt} &=&\sigma (aI_{1}+bI_{2})+\alpha _{1}t-\frac{\eta _{1}%
}{2\sqrt{I_{1}}}\cos \Phi _{1},  \label{204} \\
\frac{d\Phi _{2}}{dt} &=&\sigma (bI_{1}+cI_{2})+\alpha _{2}t-\frac{\eta _{2}%
}{2\sqrt{I_{2}}}\cos \Phi _{2},  \label{205}
\end{eqnarray}%
where%
\begin{eqnarray}
&a &=\frac{4}{\omega _{r1}^{2}}\left( k_{1}^{2}-3\sigma U_{0}^{2}\right)
k_{1}^{2}=\frac{1}{\omega _{r1}^{2}}\left( k_{1}^{4}+3\omega
_{r1}^{2}\right),   \nonumber \\
&b &=\frac{2k_{1}^{2}k_{2}^{2}}{\omega _{r1}\omega _{r2}},  \label{206}  \\
&c &=\frac{4}{\omega _{r2}^{2}}\left( k_{2}^{2}-3\sigma U_{0}^{2}\right)
k_{2}^{2}=\frac{1}{\omega _{r2}^{2}}\left( k_{2}^{4}+3\omega
_{r2}^{2}\right), \nonumber \\
&\eta _{1,2}&=\frac{\varepsilon _{1,2}|k_{1,2}|U_{0}}{\sqrt{\omega _{r1,2}}}. 
\nonumber
\end{eqnarray}%
The action-angle Hamiltonian of this system is \begin{eqnarray}
&&H(I_{1,2},\Phi_{1,2},t)=\sigma(\frac{a}{2}I_1^2+bI_1I_2+\frac{c}{2}I_2^2)\\&&+(\alpha_1I_1+\alpha_2I_2)t-\eta_1\sqrt{I_1}\cos\Phi_1-\eta_2\sqrt{I_2}\cos\Phi_2.\nonumber
\end{eqnarray}
Note that as mentioned at the end of Sec. III, in the case of two-phase driving via the modulation of the interaction strength, the system of equations (\ref{202})-(\ref{205}) remains the same, but in the expression for $\eta_{1,2}$, $U_0$ must be replaced by $2\sigma U_0^3$. Finally, a similar system of equations was derived recently in application to Langmuir and ion
acoustic waves in plasmas \cite{Munirov1,Munirov2}.

The autoresonant evolution corresponds to double phase-locking in the system as phase mismatches\ $\Phi _{1,2}$ remain bounded continuously subject
to small initial conditions on $I_{1,2}$ at large negative time $t_{0}$. In
the initial stage, the pairs of variables $(I_{1},\Phi _{1})$ and $%
(I_{2},\Phi _{2})$ decouple and are described by
\begin{equation}
\frac{dI_{i}}{dt}=-\eta _{i}\sqrt{I_{i}}\sin \Phi _{i}  \label{207}
\end{equation}%
\begin{equation}
\frac{d\Phi _{i}}{dt}=\alpha _{i}t-\frac{\eta _{i}}{2\sqrt{I_{i}}}\cos \Phi
_{i}  \label{208}
\end{equation}%
The phase-locking $\frac{d\Phi _{i}}{dt}\approx 0$ in each of these
decoupled systems is guaranteed at \textit{large negative} times (see Ref. \cite{Lazar70}
for a detailed analysis) and yields
\begin{equation}
\alpha _{1}t-\frac{\eta _{1}}{2\sqrt{I_{1}}}\cos \Phi _{1}\approx
0,\alpha _{2}t-\frac{\eta _{2}}{2\sqrt{I_{2}}}\cos \Phi _{1}\approx
0  \label{209}
\end{equation}%
and phase locking at either $\Phi _{1,2}\approx 0$ or $\pi $ if $\alpha
_{1,2}$ is negative or positive, respectively, and in both cases $\sqrt{%
I_{1,2}}\approx -\frac{\eta _{1,2}}{2\left\vert \alpha _{1,2}\right\vert t}.$

Next, assuming that the autoresonant phase-locking ($\frac{d\Phi _{i}}{dt}%
\approx 0$) continues as the system reaches \textit{large positive} times,
in double autoresonance, the actions $I_{1,2}$ are given by the
solution of%
\begin{equation}
\sigma (aI_{1}+bI_{2})+\alpha _{1}t\approx 0,\sigma (bI_{1}+cI_{2})+\alpha
_{2}t\approx 0.  \label{210}
\end{equation}%
Since coefficients $a,b,c$ are all positive, for having positive solutions
for $I_{1,2}$ chirp rates $\alpha _{1,2}$ must have the sign of $-\sigma $%
, i.e., can be written as $\alpha _{1,2}=-\sigma |\alpha _{1,2}|.$ This
yields autoresonant solutions varying linearly in time  
\begin{equation}
I_{1}=\frac{c|\alpha _{1}|-b|\alpha _{2}|}{D}t,I_{2}=\frac{a|\alpha
_{2}|-b|\alpha _{1}|}{D}t,  \label{211}
\end{equation}%
where 
\begin{equation}
D=ac-b^{2}=\frac{4k_{1}^{2}k_{2}^{2}}{\omega _{r1}^{2}\omega _{r2}^{2}}%
[4\left( k_{1}^{2}-3\sigma U_{0}^{2}\right) \left( k_{2}^{2}-3\sigma
U_{0}^{2}\right) -k_{1}^{2}k_{2}^{2}].  \label{212}
\end{equation}%
For positivness of both $I_{1,2}$ at large $t$ (large excitations) we
must have $c/b>|\alpha _{2}/\alpha _{1}|>b/a$ or%
\begin{equation}
\frac{2\omega _{r1}(k_{2}^{2}-3\sigma U_{0}^{2})}{k_{1}^{2}\omega _{r2}}%
>|\alpha _{2}/\alpha _{1}|>\frac{k_{2}^{2}\omega _{r1}}{2\omega
_{r2}(k_{1}^{2}-3\sigma U_{0}^{2})}  \label{213}
\end{equation}
Then $D=ac-b^{2}$ must be positive. This is obviously the case for $\sigma
=-1$, so we can always find some ratio $|\alpha _{2}/\alpha _{1}|$ in this
case for having positive $I_{1,2}$, linearly increasing in time. The case $%
\sigma =1$ is more complex. Since we still need $D>0$ for having increasing $%
I_{1,2}$ at large positive times, we must satisfy%
\begin{equation}
4\left( k_{1}^{2}-3U_{0}^{2}\right) \left( k_{2}^{2}-3U_{0}^{2}\right)
-k_{1}^{2}k_{2}^{2}>0  \label{214}
\end{equation}%
or%
\begin{equation}
\left( 1-X\right) \left( 1-Y\right) >1/4  \label{215}
\end{equation}%
where $X=3U_{0}^{2}/k_{2}^{2}$ and $Y=3U_{0}^{2}/k_{1}^{2}$.
\begin{figure}[tp]
\includegraphics[width=0.55\textwidth, height=0.2525\textheight, left]{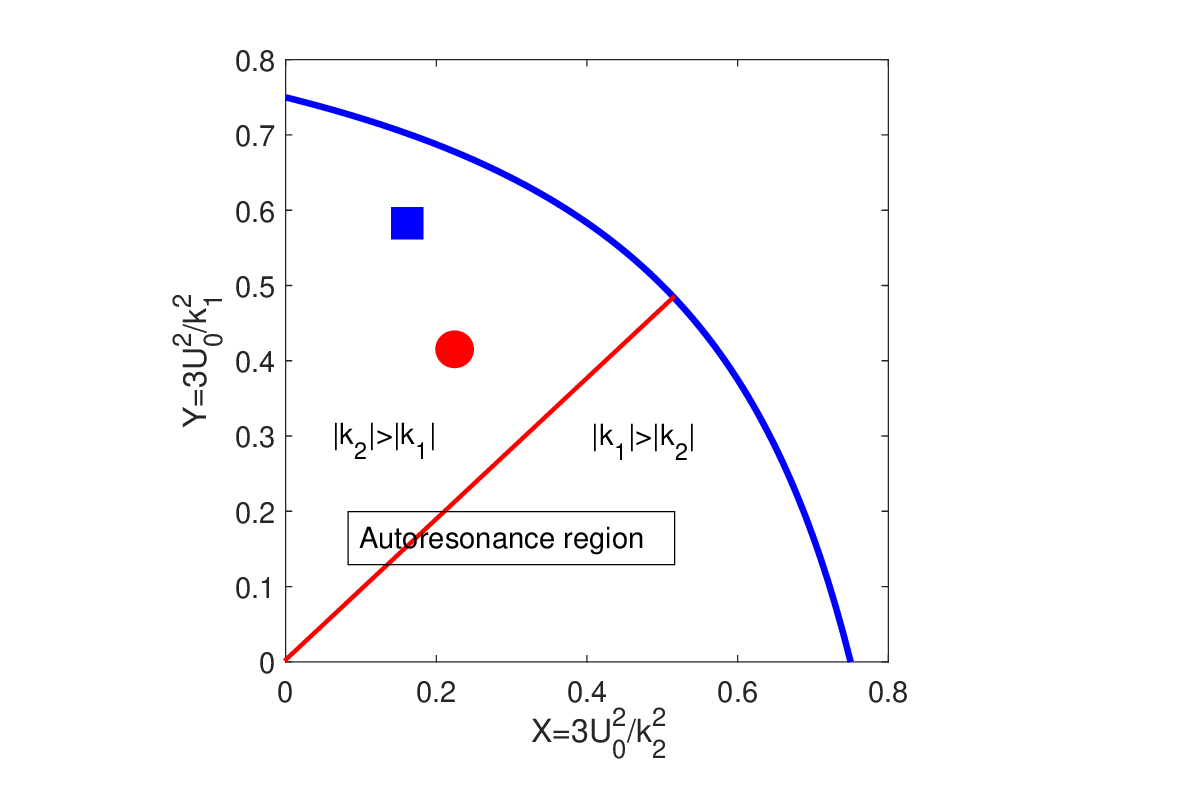}
\caption{Allowed double phase-locking region for BECs in the case $\sigma=-1$ is below the blue line in the figure. The circle and the square points correspond to parameters $k_1=2,k_2=-3,U0=0.775$ and $k_1=1,k_2=-2,U_0=0.447$, respectively. The examples of space-time quasicrystals for these  two sets of parameters are shown in Fig. 9}
\label{fig7}
\end{figure}
The last
inequality yields the condition
\begin{equation}
Y<1-\frac{1}{4(1-X)}  \label{216}
\end{equation}%
The region $S$ in the $(Y,X)$-plane satisfying this inequality (the allowed
region) is shown in Fig. 7. Note that in this region $X,Y<3/4$, which
guarantees the positivity of $\omega _{r1,2}^{2}$. However, $U_{0}^{2}$ can not be too large to satisfy $%
X,Y<3/4$ and can be chosen as follows. Let 
$k_{2}^{2}>$ $k_{1}^{2}$. Then, $X<Y$ and we can choose some value $Y_{0}<3/4
$ yielding $U_{0}^{2}=\frac{k_{1}^{2}}{3}Y_{0}$ and $X_{0}=\frac{k_{1}^{2}}{%
k_{2}^{2}}Y_{0}<\frac{k_{1}^{2}}{k_{2}^{2}}$. This guarantees that the point 
$(X_{0},Y_{0})$ is inside the allowed region if $Y_{0}<1/2$. However, if $%
1/2<Y_{0}<3/4$, for having $(X_{0},Y_{0})\in S$ we have a restriction
\begin{equation}
\frac{k_{1}^{2}}{k_{2}^{2}}<\frac{1}{Y_{0}}\left( 1-\frac{1}{4(1-Y_{0})}%
\right) .  \label{217}
\end{equation}%

The inequalities \eqref{216} and \eqref{217} above are based on the analysis at large positive times
and comprise only necesary conditions for synchronized (autoresonant)
evolution. We have already discussed the phase-locking at large negative
times. However, for synchronized passage through the vicinity of $t=0$, i.e., for
having bounded $\Phi _{1,2}$ at \textit{all times}, in addition to the
above, it requires $\eta _{1,2}$ be large enough (for both $\sigma =+1$ and  $-1
$). In dealing with this issue, we choose some value of` $\left\vert \alpha
_{1}\right\vert $, $\eta _{1}$ and $r=|\alpha _{2}/\alpha _{1}|$ ($r$ must
satisfy (\ref{213}) as described above), which defines $\left\vert \alpha
_{2}\right\vert $.
\begin{figure}[bp]
\includegraphics[width=0.51\textwidth, height=0.28\textheight, left]{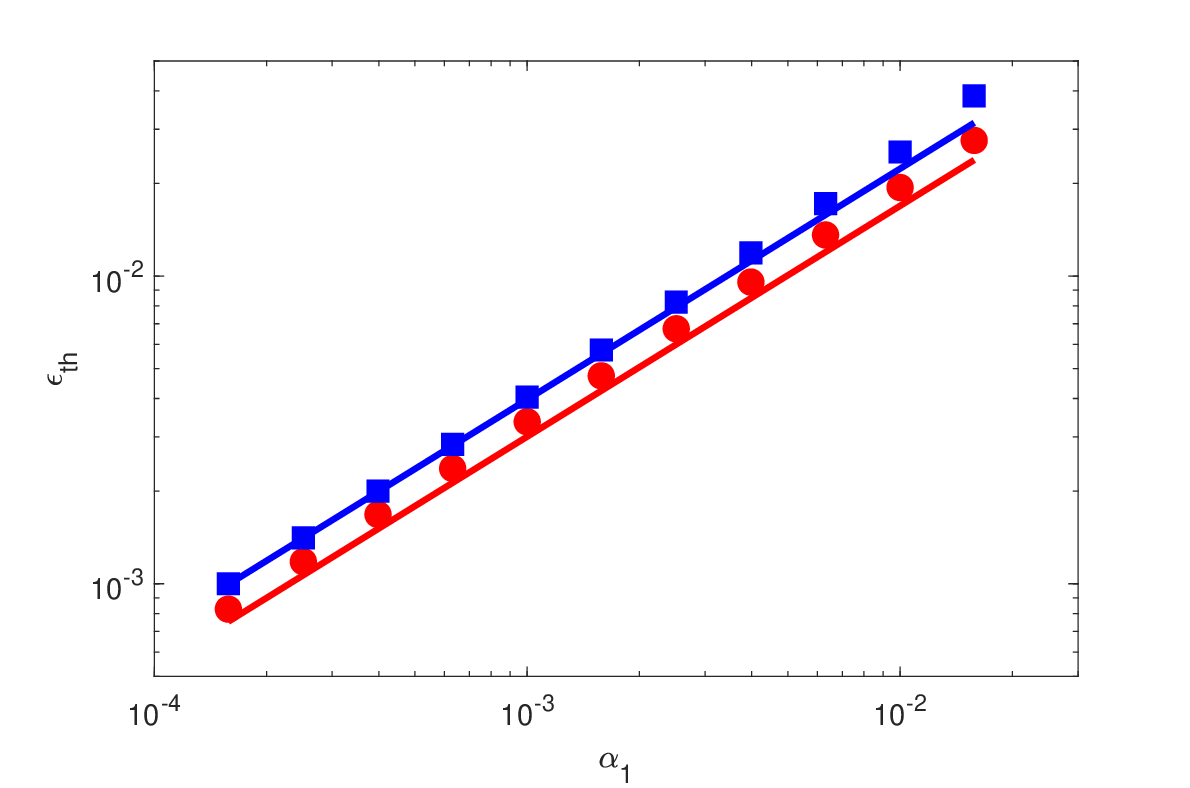}
\caption{Threshold $\varepsilon_{1th}$ versus $\alpha_1$ for autoresonant double phase-locking of BECs in the case $\sigma=-1$ and two sets of parameters, set A: $k_1=2,k_2=-3,U0=0.775,r=1.08,q=0.7,\mu_{th}=0.403$ (red line) and set B: $k_1=1,k_2=-2,U_0=0.447,r=0.805,q=1.3,\mu_{th}=0.277$ (blue line). The circles and squares show the results of full numerical simulations.}
\label{fig8}
\end{figure} 
We also fix ratio $q=\eta _{2}/\eta _{1}$, which
defines $\eta _{2}$, and we are left with the problem of finding the critical
value of $\eta _{1th}$ for autoresonant transition through the vicinity of $
t=0$ for this $\left\vert \alpha _{1}\right\vert $. Now we return to our
original system (\ref{202})-(\ref{205}) and rewrite it as\ 
\begin{eqnarray}
\frac{dJ_{1}}{d\tau } &=&-\mu \sqrt{J_{1}}\sin \Phi _{1}  \label{218} \\
\frac{dJ_{2}}{d\tau} &=&-q\mu \sqrt{J_{2}}\sin \Phi _{2}  \label{219}
\end{eqnarray}
\begin{eqnarray}
\frac{d\Phi _{1}}{d\tau} &=&\sigma \lbrack (aJ_{1}+bJ_{2})-\tau ]-\frac{\mu }{2%
\sqrt{J_{1}}}\cos \Phi _{1}  \label{220} \\
\frac{d\Phi _{2}}{d\tau } &=&\sigma \lbrack (bJ_{1}+cJ_{2})-r\tau ]-\frac{%
q\mu }{2\sqrt{J_{2}}}\cos \Phi _{2}  \label{221}
\end{eqnarray}%
where \textit{slow} time $\tau =\sqrt{\left\vert \alpha _{1}\right\vert }t$, $%
J_{1,2}=I_{1,2}/\sqrt{\left\vert \alpha _{1}\right\vert }$ and $\mu =\eta
_{1}/\left\vert \alpha _{1}\right\vert ^{3/4}$.
\begin{figure}[tp]
\includegraphics[width=0.51\textwidth, height=0.28\textheight, left]{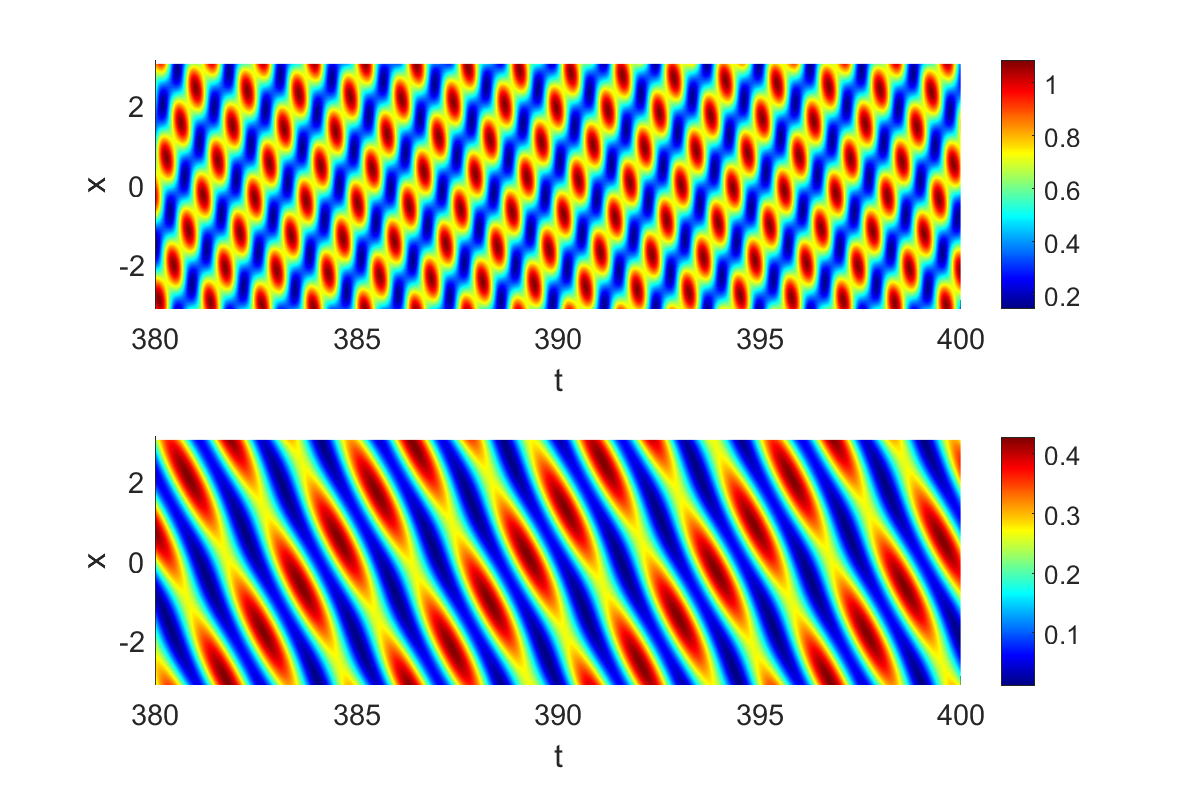}
\caption{Colormap of space-time BEC  quasicrystals for the same two sets of parameters as in Fig. 8, upper panel: set A, lower panel: set B. In both cases $\alpha_1=0.001$ and the driving amplitudes are 10\% above the threshold.}
\label{fig9}
\end{figure}
Therefore, for a given $r$ and $q$,
we are left with a single additional parameter $\mu $, and there may exists
some minimal value of $\mu _{th}$ in the problem which still guarantees a
continuous phase-locking in the system as it passes from large negative
times through $t=0$, to large positive times. This value can be found
numerically, yielding the minimal (threshold) driving amplitude $\varepsilon
_{1th}$ for autoresonance in the system:
\begin{equation}
\varepsilon _{1th}=\mu _{th}\frac{\sqrt{\omega _{r1}}}{\left\vert k_{1}\right\vert U_{0}}\left\vert
\alpha _{1}\right\vert ^{3/4}  \label{222}
\end{equation}
In the case of driving via modulation strength, the threshold formula remains the same, but $U_0$ is replaced by $2U_0^3$. 

We illustrate the characteristic $3/4$ power scaling of $\varepsilon _{1th}$
with $\left\vert \alpha _{1}\right\vert $  in Fig. 8, for the parametric driving case, $\sigma=-1$, and two sets of parameters: Set A: $k_1=2,k_2=-3,U0=0.775,r=1.08,q=0.7$ (red line) and Set B: $k_1=1,k_2=-2,U_0=0.447,r=0.805,q=1.3$ (blue line). The values of $k_{1,2}$ and $U_0$ in these two examples correspond to the two points in the allowed autoresonance region in Fig. 7. The circles and squares in Fig. 8 show the results of numerical simulations. Finally, Fig. 10 shows colormaps of $|\varphi|^2$ in space-time autoresonant space-time quasicrystals for the same two sets of parameters (the upper and lower panels correspond to sets A and B, respectively). In both cases, $\alpha_1=0.001$, and the driving amplitudes are 10\% above the thresholds in Fig. 8.

\section{Conclusions}
We have shown that a combination of two independent, small amplitude and chirped frequency parametric-type drivings or modulations of the interaction strength in the GP equation allows controlled nonlinear two-phase excitation of BECs via the process of autoresonance. The amplitude of these excitations grows continuously as the phases of the excited solution follow those of the driving perturbations. The phase-locked excitation process starts from trivial ground state at large negative times and continues as the system passes linear resonances with both driving components at $t=0$ and moves to large positive times. If both driving components are switched off at some large positive time, an ideal, stable space-time quasicrystal is formed, which is periodic in space and aperiodic in time, but preserves long time ordering (see examples of numerical simulations in Figs. 2,4, and 9). 

We have developed a weakly nonlinear theory for these two-phase periodic  excitations using Whitham's averaged variational principle, yielding a two degrees of freedom dynamical problem in action-angle variables (see Eqs. \eqref{202}-\eqref{205}). The analysis of this system at large positive times limits the parameter space for autoresonant excitations (see inequalities \eqref{216} and \eqref{217}. However, these are only necessary conditions for autoresonance in the system. A continuing phase-locking by passage through the linear resonance near $t=0$ requires the driving amplitude to surpass a certain threshold, yielding the sufficient condition for autoresonant excitation. These thresholds scale with
the driving frequency chirp rate $\alpha$ as $\sim |\alpha |^{3/4}$ (see Eq.(\ref{222})), a relationship corroborated by numerical simulations (see Fig. 8). 

Following concepts explored in this study, it seems interesting to investigate autoresonant excitation involving more complex multi-phase quasi-crystalline structures by simultaneously traversing linear resonances with three and more parametric drivings. Furthermore,
given that numerous integrable nonlinear PDEs (such as KDV, Sine-Gordon, and more) describing various physical systems allow multiphase solutions, investigating  autoresonant formation of space-time quasicrystals in these additional systems through two or more drivings using a similar approach seems interesting. Finally,  
previous investigations \cite{Fajans,Ido} have demonstrated that a sufficiently
small dissipation in other applications did not destroy single-phase autoresonant
synchronization, but modified the threshold for transition to autoresonance.
Investigating the effects of dissipation on autoresonant two-phase BECs
is also an important goal for future research. 
\section*{Acknowledgement}
This work was supported by the US-Israel Binational Science Foundation Grant No. 2020233.
\begin{widetext}
\section*{Appendix: Reduction to two degrees of freedom}
We proceed from the Lagrangian density \eqref{104} for the ponderomotive drive case
\begin{equation}
L=\frac{1}{2}\left[ U_{x}^{2}+U^{2}(V_{x}^{2}+V_{t})\right] -\frac{\sigma }{2%
}U^{4}-\frac{U^{2}}{2}[\varepsilon _{1}\cos (\theta _{1}-\Phi
_{1})-\varepsilon _{2}\cos (\theta _{1}-\Phi _{1})].  \label{A1}
\end{equation}%
The case of driving via the modulation of the interaction strength will be discussed at the end of the Appendix. We average the Lagrangian over $\theta _{1}$ and $\theta _{2}$ between $0$ and $2\pi $
using the weakly nonlinear ansatz \eqref{111} and \eqref{112}. The resulting averaged
Lagrangian density $\Lambda $ consists of the following five terms%
\begin{eqnarray}
\Lambda _{1} &=&\left\langle \frac{1}{2}U_{x}^{2}\right\rangle =\frac{1}{4}%
(k_{1}^{2}U_{1}^{2}+k_{2}^{2}U_{2}^{2})+\frac{1}{4}%
[2k_{1}k_{2}(u_{12m}^{2}-u_{12p}^{2})  \label{A2} \\
&&+k_{1}^{2}(4u_{11}^{2}+u_{12m}^{2}+u_{12p}^{2})+k_{2}^{2}(u_{12m}^{2}+u_{12p}^{2}+4u_{22}^{2})],
\nonumber
\end{eqnarray}
\begin{eqnarray}
\Lambda _{2} &=&\left\langle \frac{1}{2}U_{x}^{2}\right\rangle =\frac{%
U_{0}^{2}}{4}(k_{1}^{2}V_{1}^{2}+k_{2}^{2}V_{2}^{2})+\frac{1}{16}%
[k_{1}^{2}(8u_{0}U_{0}V_{1}^{2}+3U_{1}^{2}V_{1}^{2}  \nonumber \\
&&+4U_{0}u_{11}V_{1}^{2}+2U_{2}^{2}V_{1}^{2}+16U_{0}U_{1}V_{1}v_{11}+16U_{0}^{2}v_{11}^{2}-
\nonumber \\
&&8U_{0}(U_{2}V_{1}(v_{12m}-v_{12p})+(u_{12m}V_{1}+u_{12p}V_{1}+U_{1}(v_{12m}+v_{12p}))V_{2}))
\label{A3} \\
&&+k_{2}^{2}((2U_{1}^{2}+3U_{2}^{2})V_{2}^{2}+4U_{0}V_{2}(-2U_{1}v_{12m}+2U_{1}v_{12p}+2u_{0}V_{2}+u_{22}V_{2})
\nonumber \\
&&+4U_{0}^{2}(v_{12m}^{2}+v_{12p}^{2}+4v_{22}^{2}))],  \nonumber
\end{eqnarray}

\begin{eqnarray}
\Lambda _{3} &=&\left\langle \frac{1}{2}U^{2}V_{t}\right\rangle =\sigma
U_{0}^{4}+\frac{1}{4}(2\gamma U_{0}^{2}+2U_{0}(\sigma
U_{0}(4u_{0}U_{0}+U_{1}^{2}+U_{2}^{2})-\omega _{1}U_{1}V_{1}-\omega
_{2}U_{2}V_{2}))+  \nonumber \\
&&\frac{1}{4}[\gamma (4u_{0}U_{0}+U_{1}^{2}+U_{2}^{2})+2\sigma
U_{0}^{2}(2u_{0}^{2}+u_{11}^{2}+u_{12m}^{2}+u_{12p}^{2}+u_{22}^{2})-\omega
_{1}(2u_{0}+u_{11})U_{1}V_{1}+  \nonumber \\
&&(u_{12m}+u_{12p})U_{2}V_{1}-U_{1}^{2}v_{11}-4U_{0}u_{11}v_{11}-2U_{0}u_{12m}v_{12m}-2U_{0}u_{12p}v_{12p}+U_{1}U_{2}(v_{12m}+v_{12p}))+
\label{A4} \\
&&\omega
_{2}(U_{1}U_{2}(v_{12m}-v_{12p})+(u_{12m}+u_{12p})U_{1}V_{2}-(2u_{0}+u_{22})U_{2}V_{2}+U_{2}v_{22})+
\nonumber \\
&&U_{0}(-2u_{12m}v_{12m}+2u_{12p}v_{12p}+4u_{22}v_{22}))],  \nonumber
\end{eqnarray}

\begin{eqnarray}
\Lambda _{4} &=&\left\langle -\frac{\sigma }{2}U^{4}\right\rangle =\frac{%
\sigma }{2}U_{0}^{4}-\frac{\sigma }{2}%
U_{0}^{2}(4u_{0}U_{0}+3(U_{1}^{2}+U_{2}^{2}))-  \nonumber \\
&&\frac{3\sigma }{16}%
[16u_{0}^{2}U_{0}^{2}+U_{1}^{4}-16U_{0}U_{1}(u_{12m}+u_{12p})U_{2}+U_{2}^{4}+16u_{0}U_{0}(U_{1}^{2}+U_{2}^{2})+
\label{A5} \\
&&4U_{1}^{2}(2U_{0}u_{11}+U_{2}^{2})+8U_{0}(U_{2}^{2}u22+U_{0}(u_{11}^{2}+u_{12m}^{2}+u_{12p}^{2}+u_{22}^{2}))],
\nonumber
\end{eqnarray}

\begin{equation}
\Lambda _{5}=\left\langle -\frac{U^{2}}{2}[\varepsilon _{1}\cos (\theta
_{1}-\Phi _{1})-\varepsilon _{2}\cos (\theta _{1}-\Phi _{1})]\right\rangle =-%
\frac{U_{0}}{2}(\varepsilon _{1}U_{1}\cos \Phi _{1}+\varepsilon
_{2}U_{2}\cos \Phi _{2}),  \label{A6}
\end{equation}%
where we have expanded to 4-th order in amplitudes in Eqs. (\ref{A2})-(\ref{A5}%
) and to first-order in Eq. (\ref{A6}), assuming $\varepsilon _{1,2}$ are
sufficiently small. Note that except in $\Lambda _{5}$, the averaged
Lagrangian density $\Lambda $ includes only second and fourth-order terms
in the square brackets. Also note that $\Lambda $ does not include the time
derivatives with respect to the amplitudes, and therefore, the variations with
respect to each of these $13$ amplitudes are simply $\partial \Lambda
/\partial A_{i}=0$, where $A_{i}$ is the set of these amplitudes.

As the next step, we consider the linearized problem, i.e., neglect all
fourth-order terms in $\Lambda$. Then the variations with respect to $U_{1,2}$ and $V_{1,2}$ yield
\begin{eqnarray}
(k_{i}^{2}-4\sigma U_{0}^{2})U_{i}-\omega _{i}U_{0}V_{i}-\varepsilon
_{i}U_{0}\cos \Phi _{i} &=&0  \label{A7} \\
k_{i}^{2}U_{0}V_{i}-\omega _{i}U_{i} &=&0  \label{A8}
\end{eqnarray}%
with solutions
\begin{equation}
V_{i}=\frac{\varepsilon \omega _{i}}{(\omega _{0i}^{2}-\omega _{i}^{2})}%
\cos \Phi _{i}, U_{i}=\frac{k_{i}^{2}U_{0}}{\omega _{i}}V_{i}
\end{equation}
which is a generalization of\ Eqs. (\ref{109}) and (\ref{110}) for the
case of ideal phase locking, ($\omega_{i}=\omega _{di}$ and $\Phi_{i}=0$).

The next step is the inclusion of nonlinearities and taking variations of the
full averaged Lagrangian density $\Lambda$ with respect to $u_{0}$ and $\xi $
yielding
\begin{equation}
\frac{\partial \Lambda }{\partial u_{0}} =\gamma U_{0}-\sigma
(4u_{0}U_{0}^{2}+3U_{0}(U_{1}^{2}+U_{2}^{2}))-\frac{1}{2}(\omega
_{1}U_{1}V_{1}-k_{1}^{2}U_{0}V_{1}^{2}+\omega
_{2}U_{2}V_{2}-k_{2}^{2}U_{0}V_{2}^{2})=0,  \label{A9}
\end{equation}
\begin{equation}
\frac{d}{dt}\left( \frac{\partial \Lambda }{\partial \gamma }\right)=
\frac{d}{dt}(4u_{0}U_{0}+U_{1}^{2}+U_{2}^{2})=0.  \label{A10}
\end{equation}%
Then%
\begin{eqnarray}
u_{0} &=&-\frac{(U_{1}^{2}+U_{2}^{2})}{4U_{0}},  \label{A11} \\
\gamma &=&2\sigma (U_{1}^{2}+U_{2}^{2})+\frac{1}{2}(\omega _{1}\frac{%
U_{1}V_{1}}{U_{0}}-k_{1}^{2}V_{1}^{2}+\omega _{2}\frac{U_{2}V_{2}}{U_{0}}%
-k_{2}^{2}V_{2}^{2}).  \label{A12}
\end{eqnarray}%
In these second order results, assuming proximity to the
linear resonances, we replace $\omega_{1,2}$ with the linear resonance frequencies $%
\omega _{01,02}$ (see Eq. (\ref{110a})) and $V_{i}$ with its linear relation $%
V_{0i}=\frac{\omega _{i}}{k_{i}^{2}U_{0}}U_{i}$. Then, the term in the
brackets in the last equation vanishes and one gets:
\begin{equation}
\gamma =2\sigma (U_{1}^{2}+U_{2}^{2}).  \label{A13}
\end{equation}

Next, we take variations with respect to the remaining second-order
amplitudes, make the same replacements for $\omega _{1,2}$ and $V_{i}$ as
above and obtain
\begin{equation}
u_{ii}=-\frac{k_{i}^{2}+8\sigma U_{0}^{2}}{4U_{0}k_{i}^{2}}U_{i}^{2},
\label{A14}
\end{equation}%
\begin{equation}
v_{ii}=-\frac{\sqrt{k_{i}^{2}-4\sigma U_{0}^{2}}(k_{i}^{2}-2\sigma U_{0}^{2})%
}{2k_{i}^{3}U_{0}^{2}}U_{i}^{2},  \label{A15}
\end{equation}
\begin{equation}
u_{12p}=\frac{(4\sigma U_{0}^{2}+\sqrt{k_{1}^{2}-4\sigma U_{0}^{2}})\sqrt{%
k_{1}^{2}-4\sigma U_{0}^{2}}U_{1}U_{2}}{2k_{1}k_{2}U_{0}},u_{12m}=-u_{12p},
\label{A16}
\end{equation}
\begin{equation}
v_{1p}=\frac{k_{1}^{2}\sqrt{k_{2}^{2}-4\sigma U_{0}^{2}}-k_{2}^{2}\sqrt{%
k_{1}^{2}-4\sigma U_{0}^{2}}+(k_{1}k_{2}+4\sigma U_{0}^{2})(\sqrt{%
k_{1}^{2}-4\sigma U_{0}^{2}}-\sqrt{k_{2}^{2}-4\sigma U_{0}^{2}})}{%
2k_{1}k_{2}(k_{1}-k_{2})U_{0}^{2}}U_{1}U_{2},  \label{A17}
\end{equation}
\begin{equation}
v_{1m}=\frac{-k_{1}^{2}\sqrt{k_{2}^{2}-4\sigma U_{0}^{2}}+k_{2}^{2}\sqrt{%
k_{1}^{2}-4\sigma U_{0}^{2}}+(k_{1}k_{2}-4\sigma U_{0}^{2})(\sqrt{%
k_{1}^{2}-4\sigma U_{0}^{2}}-\sqrt{k_{2}^{2}-4\sigma U_{0}^{2}})}{%
2k_{1}k_{2}(k_{1}-k_{2})U_{0}^{2}}U_{1}U_{2}.  \label{A18}
\end{equation}%
At this stage, we take variations with respect to $V_{i}$, solve the
resulting algebraic equation for $V_{i}$, and replace again $\omega _{1,2}$ and $V_{i}$ in the nonlinear
part of the solution by $\omega _{01,02}$ and $V_{0i}=\frac{\omega _{i}%
}{k_{i}^{2}U_{0}}U_{i}$. This results in:
\begin{equation}
V_{1}=\frac{k_{1}\omega _{1}U_{0}^{2}U_{1}-(3\sigma
U_{0}^{2}U_{1}^{3}-k_{1}^{2}U_{1}(\frac{5}{8}U_{1}^{2}+\frac{3}{4}U_{2}^{2}))%
\sqrt{k_{1}^{2}-4\sigma U_{0}^{2}}}{k_{1}^{3}U_{0}^{3}},  \label{A19}
\end{equation}
\begin{equation}
V_{2}=\frac{k_{2}\omega _{2}U_{0}^{2}U_{2}+(3\sigma
U_{0}^{2}U_{2}^{3}-k_{2}^{2}U_{2}(\frac{3}{4}U_{1}^{2}+\frac{5}{8}U_{2}^{2}))%
\sqrt{k_{2}^{2}-4\sigma U_{0}^{2}}}{k_{2}^{3}U_{0}^{3}}.  \label{A20}
\end{equation}%
Note that these solutions involve first-order linear parts and third-order
nonlinear corrections.

Finally, we take variation with respect to $U_{i}$ to
get\ 
\begin{eqnarray}
(k_{1}^{2}-8\sigma U_{0}^{2})U_{1}-\omega _{1}U_{0}V_{1}+Q_{1}-\varepsilon
_{1}U_{0}\cos \Phi _{1} &=&0,  \label{A21} \\
(k_{2}^{2}-8\sigma U_{0}^{2})U_{2}-\omega _{2}U_{0}V_{2}+Q_{2}-\varepsilon
_{2}U_{0}\cos \Phi _{2} &=&0,  \label{A22}
\end{eqnarray}%
where $Q_{i}$ are third-order nonlinear corrections (similar to those in Eqs. \eqref{A19},\eqref{A20}). The last two equations can be simplified as follows: In
the nonlinear part of these equations, we again replace $\omega _{1,2}$ and $V_{i}$ by $%
\omega _{01,2}$ and $V_{0i}=\frac{\omega _{i}}{k_{i}^{2}U_{0}}%
U_{i}$. Additionally, we replace $V_{i}$ in the linear parts of (\ref{A21}) and (\ref{A22}%
) by the expressions in (\ref{A19}),(\ref{A20}). The algebra involved in these
manipulations is done via Mathematica package yielding two equations:
\begin{eqnarray}
(\omega _{01}^{2}-\omega
_{1}^{2})U_{1}-24U_{0}^{2}U_{1}^{3}+4k_{1}^{2}\sigma
U_{1}(2U_{1}^{2}+U_{2}^{2})-\varepsilon _{1}U_{0}k_{1}^{2}\cos \Phi _{1}
&=&0.  \label{A23} \\
(\omega _{02}^{2}-\omega
_{2}^{2})U_{2}-24U_{0}^{2}U_{2}^{3}+4k_{2}^{2}\sigma
U_{2}(2U_{2}^{2}+U_{1}^{2})-\varepsilon _{2}U_{0}k_{2}^{2}\cos \Phi _{2}
&=&0.  \label{A24}
\end{eqnarray}%

Finally, we need additional two equations for our reduced system describing $%
U_{i}$ and $\Phi _{i}$. These equations are obtained by variation with
respect to $\theta _{i}$. Since $\Phi _{i}(t)=\theta _{i}-\theta _{di}$, and 
$d\theta _{i}/dt=\omega _{i}$, we get%
\begin{equation}
\frac{d}{dt}\frac{\partial \Lambda }{\partial \omega _{i}}=\frac{\partial
\Lambda }{\partial \Phi _{i}},  \label{25}
\end{equation}%
which to lowest significant order yields%
\begin{equation}
\frac{dU_{i}}{dt}=-\frac{\varepsilon _{i}U_{0}k_{i}^{2}}{2\omega _{0i}}\sin
\Phi _{i}.  \label{A26}
\end{equation}

We conclude this Appendix by discussing the case of driving via the interaction strength in the GP equation.
In this case, the driving term in the Laplacian  \eqref{A1} changes (see the discussion at the beginning of Sec. II) and the Laplacian becomes \begin{equation}
L=\frac{1}{2}\left[ U_{x}^{2}+U^{2}(V_{x}^{2}+V_{t})\right] -\frac{\sigma }{2%
}U^{4}-\frac{\sigma U^{4}}{2}[\varepsilon _{1}\cos (\theta _{1}-\Phi
_{1})-\varepsilon _{2}\cos (\theta _{1}-\Phi _{1})].  \label{B1}
\end{equation}%
This change affects only the driving term $\Lambda_5$ (see Eq. \eqref{A6} in the averaged Laplacian, transforming it to
\begin{equation}
\Lambda _{5}=-%
\sigma U_{0}^3(\varepsilon _{1}U_{1}\cos \Phi _{1}+\varepsilon
_{2}U_{2}\cos \Phi _{2}).  \label{B2}
\end{equation} 
Formally, this is a replacement of $U_0$ in the coefficient from the ponderomotive drive case by $2\sigma U_0^3$. The same replacement should be done in the case of driving via interaction strength in all driving terms in the reduced system of equations \eqref{A23},\eqref{A24},\eqref{A26}.
\end{widetext}


\begin{thebibliography}{99}
\bibitem{Shechtman} D. Shechtman, I. Blech, D. Gratias, and J. W. Cahn, Phys. Rev. Lett. \textbf{53}, 1951 (1984).

\bibitem{Levine} D. Levine, P. J. Steinhardt, Phys. Rev. Lett. 
\textbf{53}, 2477 (1984).

\bibitem{Senechal} M. Senechal, \textit{Quasicrystals and Geometry} (Cambridge University Press, Cambridge, England, 1995).

\bibitem{Autti} S. Autti, V. B. Eltsov and G. E. Volovik, Phys. Rev. Lett.\textbf{120},
215301 (2018).

\bibitem{Kreil} A. J. E. Kreil, H. Yu. Musiienko-Shmarova, S. Eggert, A. A. Serga, B. Hillebrands, D. A. Bozhko, A. Pomyalov, and V. S. L'vov, Phys. Rev. B \textbf{100}, 020406(R) (2019).

\bibitem{Giergiel} K. Giergiel, A. Kuros, and K. Sacha, Phys. Rev. B \textbf{99}, 220303(R) (2019).

\bibitem{Cosme} J. G. Cosme, J. Skulte and L. Mathey, Phys Rev A \textbf{100}, 053615 (2019).

\bibitem{Scott} A. Scott, \textit{Nonlinear Science: Emergence and Dynamics of Coherent Structures} (Oxford University Press, New York,1999).

\bibitem{Novikov} S. Novikov, S.V. Manakov, L. P. Pitaevskii, and V.E. Zacharov, \textit{Theory of Solitons }(Consultants Bureau, New York,1984).

\bibitem{LazarWiki} L. Friedland, Scholarpedia \textbf{4}(1), 5473 (2009).

\bibitem{Lazar2003} L. Friedland, A.G. Shagalov, Phys. Rev. Lett. \textbf{90}, 074101 (2003).

\bibitem{Lazar2005} L. Friedland, A. G. Shagalov, Phys. Rev. E \textbf{71}, 036206 (2005).

\bibitem{Lazar2009} A. G. Shagalov, L. Friedland, Physica D \textbf{238}, 1561 (2009).

\bibitem{Munirov1} V. R. Munirov, L. Friedland, and J.S. Wurtele, Phys. Rev. E \textbf{106}, 055201 (2022).

\bibitem{Munirov2} V. R. Munirov, L. Friedland, and J.S. Wurtele, Phys. Rev. Research 4, 023150 (2022).

\bibitem{Whitham} G. B. Whitham, \textit{Linear and Nonlinear Waves} (John Wiley, NewYork, 1973).

\bibitem{Dalfovo} F. Dalfovo, S. Giorgini, L.P. Pitaevskii, S. Stringari, Rev. Mod. Phys., \textbf{71}, 463 (1999).

\bibitem{Yamazaki} R. Yamazaki, S. Taie, S. Sugawa, Y. Takahashi, Phys. Rev. Lett. \textbf{105}, 050405 (2010).

\bibitem{Schloss} J. Schloss, P. Barnett, R. Sachdeva, T. Busch, Phys. Rev A \textbf{102}, 043325 (2020).

\bibitem{Zhu} C-X. Zhu, W. Yi, G-C. Guo, Z-W. Zhou, Phys. Rev A \textbf{99}, 023619 (2019).

\bibitem{solitons} A. G. Shagalov and L. Friedland, Phys. Rev. E \textbf{109}, 014201 (2024).

\bibitem{Leggett} A. J. Leggett, Rev. Mod. Phys., \textbf{73}, 301 (2001).

\bibitem{Zakharov} V. Zakharov, L. Ostrovsky, Physica D: Nonlinear Phenomena \textbf{238}, 540 (2009).

\bibitem{Agrawal} G. P. Agrawal, \textit{Nonlinear Fiber Optics (4th ed.)} (Academic Press, San Diego, 2007) p. 121.

\bibitem{Lazar2022} A.G. Shagalov and L. Friedland, Phys. Rev. E, \textbf{106},024211(2022).

\bibitem{Lazar70} L. Friedland, Phys. Fluids B4, 3199 (1992).

\bibitem{Fajans} J. Fajans, E. Gilson, and L. Friedland, Phys. Plasmas \textbf{8}, 423 (2001).

\bibitem{Ido} I. Barth, L. Friedland, E. Sarid, and A.G. Shagalov, Phys. Rev. Lett.  \textbf{103}, 155001 (2009).

\end{thebibliography}
\end{document}